\documentclass[11pt,twoside]{article}
\usepackage{macro-fsut-eng}
\usepackage{graphicx}
\usepackage{epsfig,amsmath,amssymb,cite}
\usepackage[T1]{fontenc} 

\usepackage{latexsym}
\usepackage{verbatim}

\begin{document}

\vskip 1.0cm
\markboth{E. F. Eiroa and G. Figueroa Aguirre}{Thin-shell wormholes in Einstein-Born-Infeld theory}
\pagestyle{myheadings}

\vspace*{0.5cm}
\title{Thin-shell wormholes in Einstein-Born-Infeld theory}

\author{Ernesto F. Eiroa,$^{1,2}$ and Griselda Figueroa Aguirre,$^1$}
\affil{$^1$Instituto de Astronom\'{\i}a y F\'{\i}sica del Espacio, Buenos Aires, Argentina.\\ $^2$Departamento de F\'isica, FCEN-UBA, Buenos Aires, Argentina.}

\begin{abstract}
We construct spherically symmetric thin-shell wormholes with a generalized Chaplygin gas at the throat, in Born-Infeld electrodynamics coupled to Einstein gravity. We analyze their stability under radial perturbations. 
\end{abstract}

\section{Introduction}
\label{sec:introduction}

Traversable Lorentzian wormholes (Morris et al. 1988) are theoretical objects with a throat that connects two regions of the same universe or two different universes. They are characterized by being threaded by matter that violates the null energy condition and its amount can be arbitrarily small at the cost of increasing the pressure at the throat. A particular class of wormholes can be obtained by using the thin-shell formalism, i.e. by cutting and pasting two manifolds to construct a new one, with a shell at the joining surface corresponding to the throat (Poisson et al. 1995), which must fulfill the flare-out condition.

Born-Infeld electrodynamics is a non-linear theory proposed in order to avoid the infinite self energies of charged point particles arising in Maxwell theory. Born-Infeld type actions have appeared in low energy string theory, leading to an increasing interest in non-linear electrodynamics. The field equations obtained from the action of Born-Infeld electrodynamics coupled to Einstein gravity have the spherically symmetric vacuum solution (Bret\'on 2002): 
\begin{equation} 
ds^2=-\psi (r) dt^2+\psi (r)^{-1} dr^2+r^2(d\theta^2 + \sin^2\theta d\phi^2),
\label{metricaBI}
\end{equation}
\begin{equation} 
\psi (r) = 1 - \frac{2M}{r} + \frac{2}{3b^2}\left\lbrace r^2 - \sqrt{r^4 + b^2 Q^2} + \frac{\sqrt{|bQ|^3}}{r} F \left[ \arccos \left( \frac{r^2 - |bQ|}{r^2+|bQ|} \right) ,\frac{\sqrt{2}}{2} \right] \right\rbrace ,
\label{psiBI}
\end{equation}
with $M$ the mass, $Q$ the charge, and $F(\gamma ,k)$ the elliptic integral of the first kind.

\section{Wormhole construction and stability analysis}
\label{sec:tswh}

\begin{figure}[t!]
\begin{center}
\includegraphics[width=0.47\linewidth]{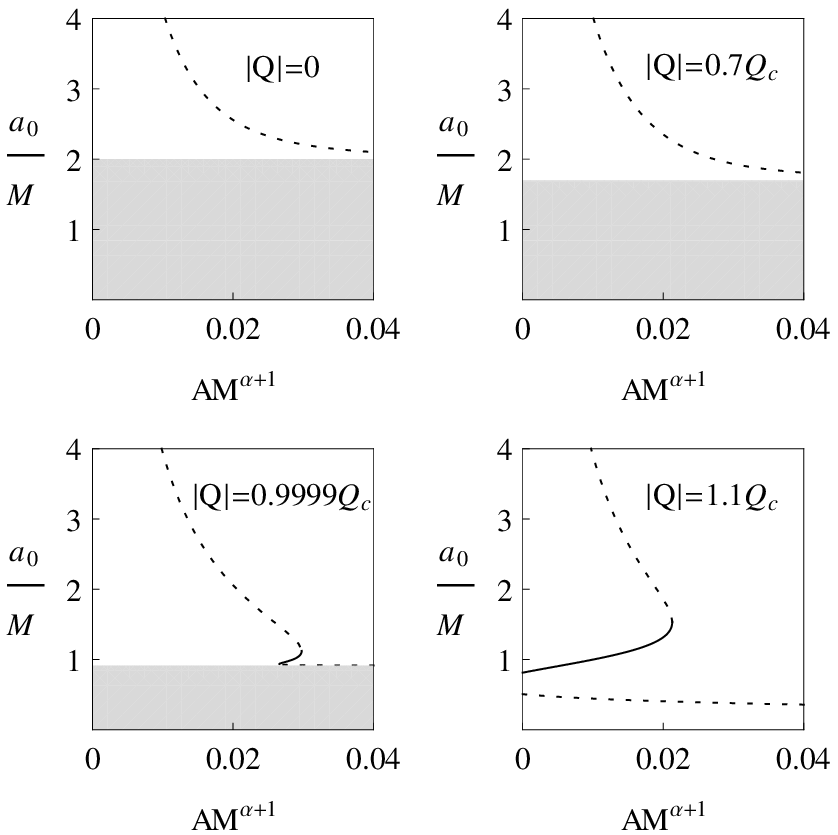}
\hspace{0.01\linewidth} \vrule{} \hspace{0.01\linewidth}
\includegraphics[width=0.47\linewidth]{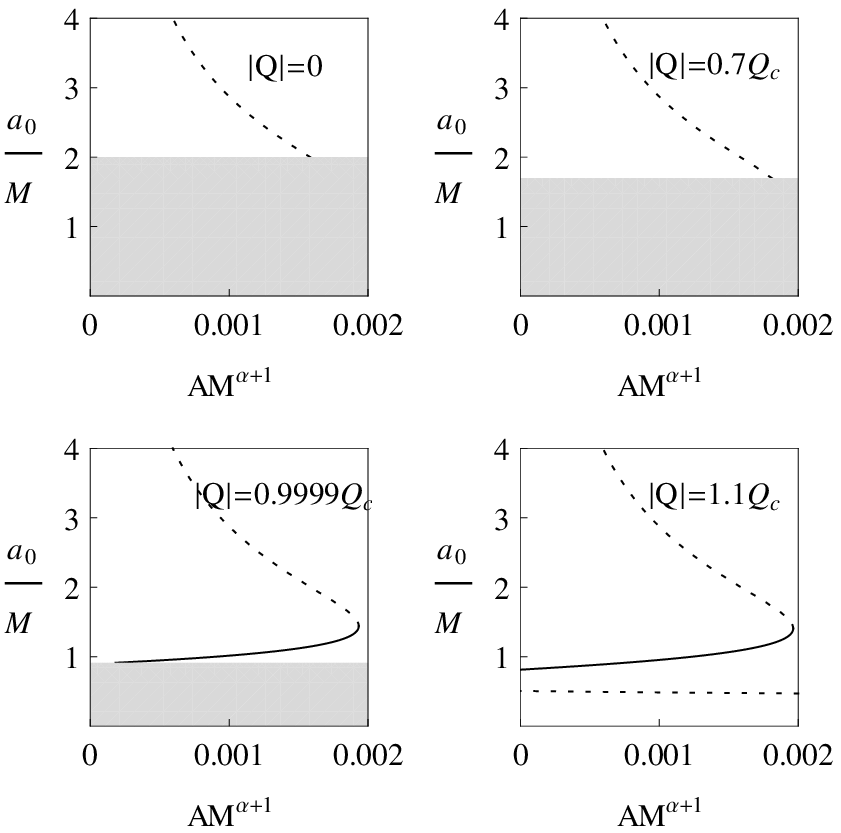}
\end{center}
\caption{Stability for $b/M=1$, in this case $Q_c/M=1.02526$. Left: $\alpha=0.2$, right: $\alpha=1$.}
\label{fb1}
\end{figure}

\begin{figure}[t!]
\begin{center}
\includegraphics[width=0.47\linewidth]{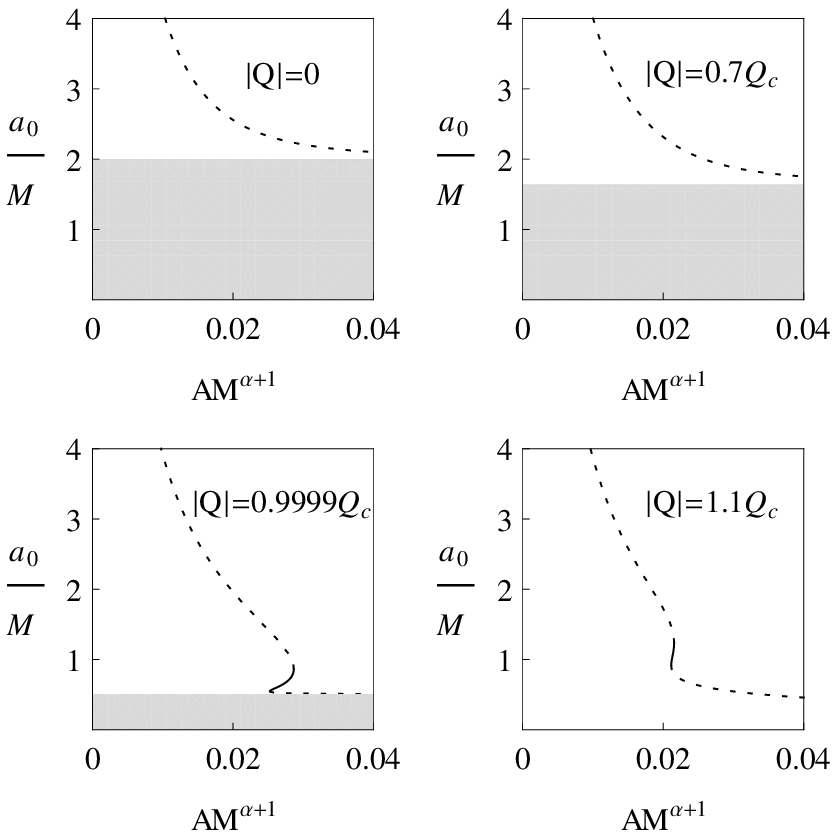}
\hspace{0.01\linewidth} \vrule{} \hspace{0.01\linewidth}
\includegraphics[width=0.47\linewidth]{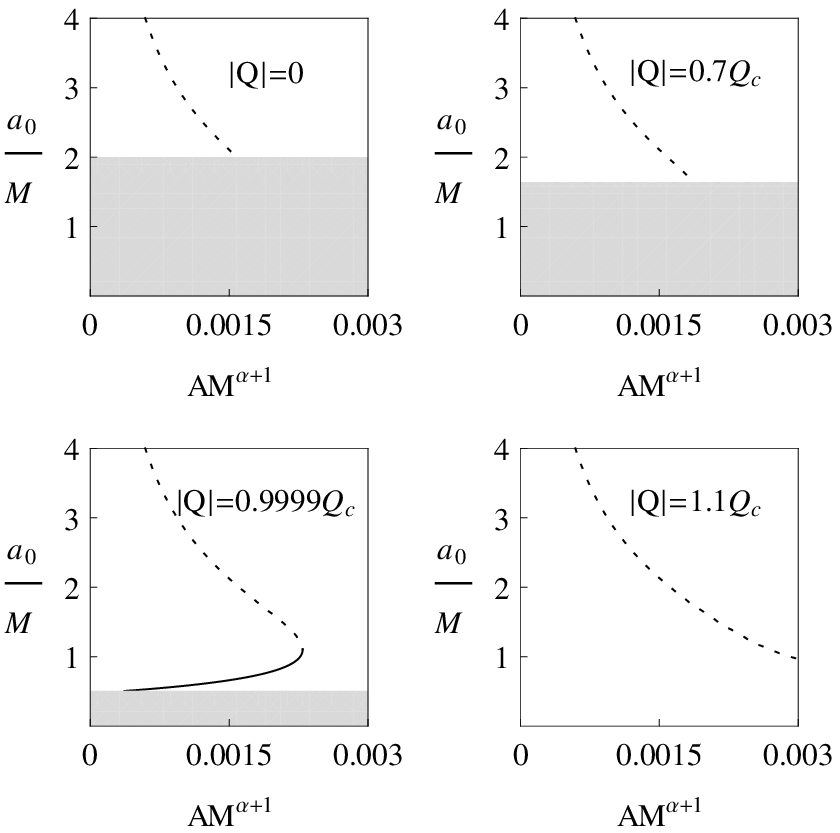}
\end{center}
\caption{Stability for $b/M=2$, in this case $Q_c/M=1.10592$. Left: $\alpha=0.2$, right: $\alpha=1$.}
\label{fb2}
\end{figure}

From the geometry (\ref{metricaBI}) we construct the thin-shell wormholes, by using the  Darmois-Israel formalism (Israel 1966). We cut and paste two identical copies of the region $r\ge a$; then at $r=a$ there is a shell where the throat is located. We let $a=a(\tau )$, with $\tau$ the proper time on the shell, and we take $a$ larger than the horizon radius $r_h$, in order to avoid the presence of the horizons and the 
singularity in the new manifold. The Einstein equations on the shell can be reduced to Lanczos equations that relate the extrinsic curvature with the surface stress-energy tensor $S_{_{\hat{\imath}\hat{\jmath} }}={\rm diag}(\sigma ,p_{\hat{\theta}},p_{\hat{\varphi}})$, with $\sigma$ the surface energy 
density and $p_{\hat{\theta}}$, $p_{\hat{\varphi}}$ the transverse pressures, from which we obtain
\begin{equation} 
\sigma=-\frac{\sqrt{\psi(a)+\dot{a}^2}}{2\pi a},
\label{e9}
\end{equation}
\begin{equation}
p=p_{\hat{\theta}}=p_{\hat{\varphi}}=\frac{\sqrt{\psi(a)+\dot{a}^2}}{8\pi} \left[\frac{2}{a} + \frac{2\ddot{a}+\psi '(a)}{\psi(a)+\dot{a}^2}\right] .
\label{e10}
\end{equation}
From Eq. (\ref{e9}) we see that $\sigma <0$, which indicates the presence of exotic matter at the junction shell. In a previous related work (Richarte et al. 2009), the matter at throat was modelled by a gas with a linearized equation of state; here we adopt a generalized Chaplygin gas (Bento et al. 2002), with an equation of state 
\begin{equation}
p=A|\sigma |^{-\alpha},
\label{e11} 
\end{equation} 
where $A>0$ and $0<\alpha\le 1$ are constants. The generalized Chaplygin gas has been adopted in current cosmology studies in order to explain the accelerated expansion of the universe. 
The dynamical evolution of the wormhole throat can be obtained by replacing Eqs. (\ref{e9}) and (\ref{e10}) into Eq. (\ref{e11}) to give
\begin{equation}
\left\{ [2\ddot{a}+\psi' (a)]a^2+[\psi(a)+\dot{a}^2]2a \right\}[2a]^{\alpha}-2A[4\pi a^{2}]^{\alpha +1}[\psi (a)+\dot{a}^2]^{(1-\alpha)/2}=0.
\label{e12} 
\end{equation}
This equation should be satisfied by thin-shell wormholes in Einstein-Born-Infeld theory, threaded by exotic matter with the equation of state of a generalized Chaplygin gas. If static solutions exist, they should satisfy Eq. (\ref{e12}) evaluated at a constant $a_{0}$. We can obtain $\sigma = \sigma (a)$ by the integration of the equation
\begin{equation}
\dot{\sigma}=-2\left( \sigma + p\right) \frac{\dot{a}}{a},
\label{p2}
\end{equation}
which comes from the the conservation equation.
Then, we replace $\sigma(a) $ in Eq. (\ref{e10}) to find  the equation that determines completely the dynamics of the throat:
\begin{equation}
\dot{a}^{2}=-V(a)=-\left\{ \psi(a)-\left[2\pi a\sigma (a)\right] ^{2}\right\},
\label{p4}
\end{equation}
where $V(a)$ can be interpreted as a potential, so it can be expanded in a Taylor series in order to analyze the stability of static solutions.
It is not difficult to see that $V(a_{0})=V'(a_{0})=0$, so the stability condition is given by $V''(a_{0})>0$, which takes the form (Eiroa et al. 2012)
\begin{equation}
V''(a_{0})=\psi ''(a_{0})+\frac{(\alpha -1) [\psi '(a_{0})]^2 }{2\psi (a_{0})} + \frac{\psi '(a_{0})}{a_{0}}- \frac{2(\alpha +1)\psi (a_{0})}{a_{0}^2}>0.
\label{p9}
\end{equation}
By using Eq. (\ref{e12}) evaluated in $a_{0}$, we can find the possible throat radii $a_{0}$, for different values of the Born-Infeld parameter $b$, the constant $A$, the exponent $\alpha$, the mass $M$ and  the charge $Q$. These solutions are stable if inequality (\ref{p9}) is fulfilled. The results shown in Figs. \ref{fb1} and \ref{fb2} present an important change around $Q_c/M$, where $Q_{c}$ is the critical charge, corresponding to the extremal value, from which the original metric used in the construction has no horizons. In the plots, the stable solutions are displayed with solid lines, the dotted lines correspond to unstable configurations; and the regions that have no physical meaning are shaded in gray. 

\begin{itemize}
\item  For $b=0$ (then $Q_c/M=1$), the Born-Infeld electrodynamics reduces to Maxwell theory, so the Reissner-Nordstr\"{o}m solution is used in the wormhole construction (Eiroa, 2009).
\item  If $0<b/M\leq1$ the behavior of the solutions is similar to what is shown in Fig. \ref{fb1}, corresponding to $b/M=1$ (then $Q_c/M=1.02526$):

Case $0<\alpha<1$ (for example, $\alpha = 0.2$): For $0\leq |Q|<Q_c$ and $|Q|$ not very close to $Q_c$, there is one unstable solution for each value of $AM^{\alpha +1}$, and this behavior continues when this parameter grows.
For $|Q|\lesssim Q_c$, we find three solutions, one of them stable; when $AM^{\alpha +1}$ grows there is only one unstable solution close to the radius of the horizon of the original manifold.
For $|Q|>Q_c$, there are three solutions, only one of them is stable.

Case $\alpha=1$: For $0\leq |Q|<Q_c$ and $|Q|$ not very close to $Q_c$, there is only one solution which is unstable. For $|Q|\lesssim Q_c$, there are two solutions, one stable and the another unstable. For $|Q|>Q_c$, there are three solutions, only one of them is stable. 

Comparing the results shown in Fig. \ref{fb1} with those obtained in Reissner-Nordstr\"om case (Eiroa, 2009), we observe a similar behavior when $b=0$ and when $0<b/M\leq 1$, for the same values of $\alpha$. The only difference is found in the case $|Q|>Q_c$, where there is a small unstable solutions for $0<b/M\leq 1$ which is not present in $b=0$ case.
\item From Fig. \ref{fb2}, corresponding to $b/M=2$ (then $Q_c/M=1.10592$):

Case $0<\alpha<1$ (for example, $\alpha = 0.2$): For $0\leq |Q|<Q_c$ and $|Q|$ not very close to $Q_c$, there is always one unstable solution.
For $|Q|\lesssim Q_c$ or $|Q|>Q_c$, there exists three solutions, one of them is stable.

Case $\alpha=1$: For $0\leq |Q|<Q_c$ or $|Q|>Q_c$, with  $|Q|$ not very close to $Q_c$, there is only one unstable solution. For $|Q|\thicksim Q_c$, there are two solutions, one of them stable.

\item When $b/M$ takes large values (not shown in the figures), for example if $b/M=5$, (then $Q_c/M=1.48468$), for any values of $\alpha $ and $Q$ there is only one solution, which is always unstable (Eiroa et al. 2012).
\end{itemize}

\section{Conclusions}
\label{sec:conclu}

For small values of $b/M$, the behavior of the solutions resemble the one obtained for Reissner-Nordstr\"om metric, except that in the Einstein-Born-Infeld case are found unstable solutions for large values of $|Q|/M$. As $b/M$ increases, i.e. when the theory is distancing itself from Einstein-Maxwell, the stability region becomes smaller. For large values of $b/M$ the stable solutions are not longer present. 

\bigskip

\acknowledgments

This work was supported by CONICET and UBA.


\begin{references}

Bento, M. C., Bertolami, O., Sen, A. A. 2002, Phys. Rev. D 66, 043507.

Bret\'{o}n, N. 2002, Class. Quantum Grav. 19, 601.

Eiroa, E. F. 2009, Phys. Rev. D, 80, 044033.

Eiroa, E. F., Figueroa Aguirre, G. 2012, Eur. Phys. J. C 72, 2240.

Israel, W. 1966, Nuovo Cimento B 44, 1; 1967, ibid. 48, 463(E).

Morris, M. S., Thorne, K. S. 1988, Am. J. Phys. 56, 395.

Poisson, E., Visser, M. 1995, Phys. Rev. D 52, 7318.

Richarte, M. G., Simeone, C. 2009, Phys. Rev. D 80, 104033; 2010, ibid. 81, 109903(E).

\end{references}
\end{document}